\documentclass[runningheads]{llncs}
\usepackage[T1]{fontenc}
\usepackage{graphicx}
\usepackage{amsmath}
\usepackage{amssymb} 
\usepackage{orcidlink}
\usepackage{caption}
\usepackage{subcaption}
\usepackage{booktabs}
\usepackage{multirow}
\makeatletter
\renewcommand\subparagraph{\@startsection{subparagraph}{5}{\z@}{3.25ex \@plus1ex \@minus.2ex}{-1em}{\normalfont\normalsize\bfseries}}
\makeatother
\usepackage{titlesec}
\usepackage{enumitem}

\titlespacing*{\subsection}{0pt}{0.5em}{1em}
\titlespacing*{\subsubsection}{0pt}{0.5em}{1em}

\begin{document}
\title{Bridging the Closed-World Gap: Interpretable Anomaly Detection via a Hybrid TSK-FCM Ensemble on the BETH Dataset}
\titlerunning{Fuzzy Ensemble Anomaly Detection}

\author{Lohith Srikanth Pentapalli\,\orcidlink{0009-0002-9801-3418} \and Josette Riep\orcidlink{0000-0002-7325-4381} \and Justin Ouwerkerk \and Kelly Cohen\,\orcidlink{0000-0002-8655-1465}}
\authorrunning{Pentapalli L.S., et al.}

\institute{University of Cincinnati \\
\email{pentapls@mail.uc.edu, riepjr@ucmail.uc.edu, ouwerkjn@ucmail.uc.edu, cohenky@ucmail.uc.edu}}

\maketitle

\begin{abstract}
A fundamental paradox plagues the deployment of Machine Learning in cybersecurity: supervised models generalize poorly to novel, Out-of-Distribution (OoD) attacks, whereas unsupervised models often suffer from poor separability, leading to unmanageable false-alarm rates. This challenge is exemplified by the BETH dataset, a kernel-level benchmark in which the training data consists solely of benign and ``suspicious'' insider behaviors, whereas the test set consists almost entirely of novel botnet activation attacks. To bridge this detection gap, we propose a Hybrid Fuzzy Ensemble that synergizes the precision of supervised learning with the adaptability of unsupervised clustering. Our architecture integrates a gradient-optimized Takagi-Sugeno-Kang (TSK) fuzzy classifier to act as an expert on known insider threats, with a parallel Fuzzy C-Means (FCM) module to detect geometric anomalies. A critical finding of our research is that traditional fuzzy membership scores degrade in high-dimensional feature spaces due to the curse of dimensionality. We overcome this by introducing a robust \textit{Min Distance} metric that exploits the raw geometric isolation of botnet activation attacks. This approach yields a state-of-the-art AUROC of \textbf{0.9951} on the BETH test set, outperforming the original Isolation Forest baseline by \textbf{17.1\%}. By combining near-perfect OoD detection with the intrinsic interpretability of fuzzy logic, our framework offers a viable path toward robust, transparent, and autonomous network defense.

\end{abstract}

\section{Introduction}
\label{sec:intro}

The evolution of cyber threats has necessitated the use of increasingly sophisticated systems to prevent and mitigate the organizational impact of cyberattacks. While Machine Learning (ML) based Intrusion Detection Systems (IDS) have shown promise, they face a fundamental dichotomy: supervised models excel at detecting known threats but fail catastrophically against zero-day exploits (closed-world assumption), while unsupervised models can detect anomalies but often suffer from prohibitive false alarm rates due to a lack of semantic context \cite{zoppiUnsupervisedAnomalyDetectors2021}.

This challenge is clearly demonstrated by the BETH dataset \cite{highnam2021bethdata}, a modern cybersecurity benchmark collected from real-world honeypots. Unlike legacy datasets derived from synthetic traffic (e.g., NSL-KDD), BETH captures raw kernel-level system calls and exhibits a strict distributional shift: the training set contains only benign traffic and ``suspicious'' insider anomalies, while the test set introduces entirely novel ``evil'' botnet activation  \cite{highnam2021bethdata}. This structure mimics the hardest reality of cybersecurity: an IDS must defend against attack vectors it has never seen before.

Standard supervised classifiers, including Deep Neural Networks, struggle with the BETH dataset because the ``evil'' class is Out-of-Distribution (OoD) relative to the training manifold \cite{highnam2021bethdata}. Conversely, purely unsupervised methods often fail to distinguish between rare benign administrative tasks and genuine malicious behavior. To bridge this gap, we propose a \textbf{Hybrid Fuzzy Ensemble} that synergizes the strengths of both paradigms.

Our architecture integrates a gradient-optimized Takagi-Sugeno-Kang (TSK) fuzzy system with a Fuzzy C-Means (FCM) anomaly detector. The TSK module acts as a supervised expert for known insider threats, providing interpretable rule-based detection. Simultaneously, the FCM module constructs a geometric manifold of normality, flagging novel botnet activation attacks based on their raw Euclidean distance from benign clusters.


Our proposed hybrid architecture yields an AUROC of \textbf{0.9951}, surpassing the state-of-the-art Isolation Forest \cite{zoppiUnsupervisedAnomalyDetectors2021} baseline by 17.1\% while maintaining high operational interpretability.

\section{Background}

Contemporary Intrusion Detection Systems (IDS) exhibit a fundamental weakness when exposed to Out-of-Distribution (OoD) data, particularly zero-day attacks that deviate from the training distribution. Supervised learning models, including Random Forests and Deep Neural Networks (DNNs), operate under a closed-world assumption and rely on rigid decision boundaries derived from labeled data. While effective for known attack signatures, these models degrade sharply when confronted with novel threats, often producing confident misclassifications due to poor generalization beyond the training manifold \cite{rakineComprehensiveReviewIntrusion2025}.

This limitation is especially pronounced in kernel-level telemetry environments such as those represented by the BETH dataset. In such settings, anomalous behavior is more accurately characterised as a distributional shift driven by unseen adversarial techniques rather than isolated statistical outliers \cite{highnam2021bethdata}. Stream-based unsupervised learning approaches have therefore been shown to be more suitable, as they adapt to concept drift and can identify emergent process behaviors, such as unexpected system call sequences, that supervised baselines fail to detect \cite{ereminUnsupervisedAnomalyDetection2025}. Among these approaches, clustering-based methods have demonstrated particular promise. Fuzzy C-Means (FCM) clustering outperforms hard partitioning techniques by allowing soft membership assignments, enabling the identification of suspicious instances that reside in sparse or ambiguous regions of the feature space. This capacity to explicitly model uncertainty makes FCM well suited for zero-day attack detection, where class labels and attack semantics are unknown \cite{sahooModifiedFuzzyCMeans2024}.

As IDS models increase in complexity, explainability has emerged as a critical operational requirement. Black-box deep learning models typically provide probability scores without semantic justification, which limits analyst trust and slows response in Security Operations Centers (SOCs) \cite{iantornoFuzzyLogicCybersecurity2025}. In contrast, Fuzzy Logic systems offer intrinsic interpretability through rule-based reasoning expressed in linguistic terms, closely approximating human expert judgment. This transparency allows analysts to trace alerts directly to the conditions and features that triggered them, improving trust and operational usability.

Recent advances have significantly reduced the performance gap between fuzzy systems and deep learning. Gradient-optimized Takagi–Sugeno–Kang (TSK) fuzzy frameworks leverage modern optimization techniques to fine-tune membership functions, achieving near-perfect classification performance in security tasks while retaining full rule-level explainability \cite{pentapalliGradientOptimizedTSKFuzzy2025}. These findings support the use of TSK models as white-box detection engines capable of delivering both high accuracy and transparent decision logic in high-stakes cybersecurity environments. Additionally, TSK fuzzy systems have been proven to be universal approximators \cite{yingGeneralTakagiSugenoFuzzy1998}.

The literature further indicates that hybrid architectures provide an effective solution to the trade-off between detection performance and false-alarm rate. Purely supervised models tend to miss novel attacks, whereas unsupervised detectors often overproduce false positives. Hybrid systems that combine multiple learning paradigms have been shown to significantly reduce false alarm rates by leveraging complementary strengths across models \cite{hariharanHybridDeepLearning2025}. In fuzzy-based systems, initializing supervised or rule-based learners with clustering-derived structure has been demonstrated to improve convergence stability and prevent suboptimal local minima during optimization \cite{decampossouzaDetectionAnomaliesLargeScale2020}. Large-scale ensemble frameworks further validate this approach, achieving extremely low false alarm rates by combining weak learners with probabilistic calibration mechanisms \cite{abouelasaadAegisGuardMultiStageHybrid2025}.

Overall, the literature supports a hybrid paradigm in which unsupervised learning discovers latent structure and emergent attack patterns in data, while supervised or rule-based components refine decision boundaries and provide interpretable explanations. In this context, combining FCM with a gradient-optimized TSK fuzzy system offers a principled approach to minimizing both missed detections and false alarms while maintaining operational transparency.

\section{Methodology}
\label{sec:methodology}

We propose a three-stage hybrid fuzzy ensemble architecture that combines supervised detection of known suspicious behaviors with unsupervised identification of out-of-distribution (OoD) malicious events. The system consists of a gradient-optimized Takagi-Sugeno-Kang (TSK) fuzzy classifier for supervised detection and a Fuzzy C-Means (FCM) clustering module for unsupervised OoD detection. This dual-module approach addresses the fundamental challenge in botnet activation detection: identifying novel attack patterns that were not present during training.

\subsection{Data Preprocessing}
\label{subsec:preprocessing}

The BETH dataset \cite{highnam2021bethdata} consists of kernel log traces captured from 23 honeypot networks during normal operation and botnet infection periods. Each system call event is represented by 14 raw features, including process identifiers, user contexts, and system call parameters.

Following the BETH benchmark methodology, we extract 9 discriminative features for our detection system:

\begin{itemize}
    \item \textbf{Process Features}: \texttt{processId}, \texttt{parentProcessId}
    \item \textbf{User Context}: \texttt{userId}, \texttt{mountNamespace}
    \item \textbf{System Call Context}: \texttt{eventId}, \texttt{argsNum}, \texttt{returnValue}
    \item \textbf{Categorical Features}: \texttt{processName}, \texttt{hostName}
\end{itemize}

To handle the heterogeneous nature of system call features, we apply domain-specific transformations:

\begin{enumerate}
    \item \textbf{Binary Encoding}: Following security best practices, process and user identifiers are binarized based on privilege levels:
    \begin{align}
        \text{processId}' &= \mathbf{1}[\text{processId} \in \{0,1,2\}] \\
        \text{userId}' &= \mathbf{1}[\text{userId} < 1000]
    \end{align}
    where system-critical processes (PIDs 0-2) and system users (UID $<$ 1000) are flagged.
    
    \item \textbf{Categorical Encoding}: Process and host names are encoded using label encoding to preserve ordinal relationships discovered during training.
    
    \item \textbf{Return Value Mapping}: System call return values are mapped to three semantic categories:
    \begin{equation}
        \text{returnValue}' = \begin{cases}
            -1 & \text{if error (} < 0\text{)} \\
            0 & \text{if success (} = 0\text{)} \\
            1 & \text{if data returned (} > 0\text{)}
        \end{cases}
    \end{equation}
    
    \item \textbf{Standardization}: All features are normalized using z-score standardization using the training data and applied consistently to validation and test data.
\end{enumerate}

\subsection{System Architecture}
\label{subsec:architecture}

\subsubsection{Stage 1: Supervised TSK Fuzzy Classifier --}

The first detection module is a gradient-optimized first-order Takagi-Sugeno-Kang fuzzy inference system trained on labeled suspicious behaviors. Building upon our prior work on explainable phishing detection \cite{pentapalliGradientOptimizedTSKFuzzy2025}, we adapt the TSK architecture to the botnet detection domain.


Each input feature $x_i$ is fuzzified using Gaussian membership functions:
\begin{equation}
    \mu_{A_{ij}}(x_i) = \exp\left(-\frac{(x_i - c_{ij})^2}{2\sigma_{ij}^2}\right)
\end{equation}
where $c_{ij}$ and $\sigma_{ij}$ are the center and width parameters of the $j$-th membership function for feature $i$. These parameters are initialized using k-means clustering on training data and subsequently optimized via gradient descent.


For each rule $k$, the firing strength $\alpha_k$ is computed using the product t-norm:
\begin{equation}
    \alpha_k = \prod_{i=1}^{n} \mu_{A_{ik}}(x_i)
\end{equation}
where $A_{ik}$ denotes the antecedent membership function selected for feature $i$ in rule $k$.

Each rule has a linear consequent function:
\begin{equation}
    f_k(\mathbf{x}) = b_k + \sum_{i=1}^{n} w_{ki} x_i
\end{equation}

The final TSK output is the normalized weighted average:
\begin{equation}
    y_{TSK} = \frac{\sum_{k=1}^{R} \alpha_k f_k(\mathbf{x})}{\sum_{k=1}^{R} \alpha_k + \epsilon}
\end{equation}
where $\epsilon = 10^{-6}$ prevents division by zero. This output is passed through a sigmoid activation function $\sigma(\cdot)$ to produce a probability score $P_{sus} \in [0, 1]$.


We formulate the entire TSK system as a differentiable computation graph. We jointly optimize all antecedent parameters $\{c, \sigma\}$ and consequent parameters $\{w, b\}$ using the Adam optimizer to minimize the Binary Cross Entropy (BCE) loss:
\begin{equation}
    \mathcal{L} = -\frac{1}{N} \sum_{i=1}^N \left[ y_i \log(P_{sus}^{(i)}) + (1-y_i) \log(1 - P_{sus}^{(i)}) \right] + \lambda \|\mathbf{w}\|_2^2
\end{equation}
The inclusion of $L_2$ regularization on the consequent weights prevents overfitting and ensures smooth decision boundaries. The optimization is performed with weighted loss to handle class imbalance, setting the positive class weight proportional to the class ratio in the training data. We used the Adam optimizer with a learning rate of 0.01 trained over 10 epochs. The Adam optimizer was chosen due to its adaptive learning rates and stable convergence properties \cite{kingmaAdamMethodStochastic2017}.

\subsubsection{Stage 2: Unsupervised FCM OoD  --}

The second module addresses the fundamental limitation of supervised approaches: detecting novel attack patterns absent from training data. We employ Fuzzy C-Means clustering trained exclusively on benign system call patterns to model the ``manifold of normality.''


Critically, the FCM module is trained only on samples labeled as non-suspicious ($\text{sus} = 0$) in the training set. This aligns with the true out-of-distribution detection paradigm where malicious samples represent genuinely novel patterns not seen during training.


The FCM algorithm partitions the benign training data $\{\mathbf{x}_1, \ldots, \mathbf{x}_N\}$ into $c$ fuzzy clusters by minimizing:
\begin{equation}
    J_m = \sum_{i=1}^{N} \sum_{j=1}^{c} u_{ij}^m \|\mathbf{x}_i - \mathbf{v}_j\|^2
\end{equation}
subject to the constraint $\sum_{j=1}^c u_{ij} = 1$ for all $i$, where:
\begin{itemize}
    \item $\mathbf{v}_j \in \mathbb{R}^n$ is the centroid of cluster $j$
    \item $u_{ij} \in [0,1]$ is the membership degree of sample $i$ to cluster $j$
    \item $m = 2$ is the fuzzification parameter controlling cluster overlap
\end{itemize}

In standard fuzzy anomaly detection, the "outlierness" of a sample is often defined as $1 - \max(u_{ij})$. However, our empirical analysis reveals that this metric underperforms in high-dimensional spaces due to the "curse of dimensionality." As the volume of the feature space expands, OoD samples often become equidistant to multiple cluster centers, resulting in low but uniform membership distributions that are indistinguishable from benign boundary points.

To overcome the limitations of membership scores, we propose a geometric anomaly metric based on raw Euclidean distance. We define the OoD score $S_{OoD}$ as the distance to the \textit{nearest} benign prototype:
\begin{equation}
    S_{OoD}(\mathbf{x}) = \min_{j \in \{1 \dots c\}} \| \mathbf{x} - \mathbf{v}_j \|_2
\end{equation}

This metric leverages the geometric concentration of measure. Benign samples, by definition, cluster tightly around the learned centroids $\mathbf{v}_j$. Malicious botnet attacks, which generate anomalous system call sequences (e.g., rapid process forking or unauthorized file access), manifest as geometric outliers far removed from any benign cluster. By measuring the raw Euclidean distance, we bypass the normalization effects of the softmax/membership function and directly quantify the geometric isolation of the attack.

\subsubsection{Stage 3: Hybrid Fuzzy Aggregation --}
The final decision logic combines the supervised probability $P_{sus}$ and the unsupervised distance score $S_{OoD}$ into a unified threat score.

Since $P_{sus} \in [0,1]$ and $S_{OoD} \in [0, \infty)$, they cannot be compared directly. We employ a data-driven calibration step to map the distance score to a probability space. We fit a Sigmoid function to the distances observed in the validation set:
\begin{equation}
    P_{OoD}(\mathbf{x}) = \frac{1}{1 + \exp(-\alpha(S_{OoD}(\mathbf{x}) - \beta))}
\end{equation}
where $\beta$ is the median distance of benign validation samples, and $\alpha$ is a scaling factor derived from the interquartile range.

The final ensemble prediction is derived using a fuzzy T-Conorm (Maximum) operator:
\begin{equation}
    Y_{Final}(\mathbf{x}) = \max(P_{sus}(\mathbf{x}), P_{OoD}(\mathbf{x}))
\end{equation}
This "Max-Dominance" strategy creates a logical \textbf{OR} gate between the two experts:
\begin{itemize}[topsep=0pt]
    \item If the sample resembles a known insider threat, $P_{sus}$ will be high ($> 0.5$), triggering a detection.
    \item If the sample is a novel botnet attack, it will be geometrically distant, causing $P_{OoD}$ to be high ($> 0.5$), triggering a detection.
\end{itemize}
This hybrid architecture ensures robust coverage of the threat landscape, allowing the system to detect specific known threats with high precision while maintaining the generalization capability required for zero-day defense. 

\section{Results and Discussion}
\label{sec:results}

In this section, we present a comprehensive evaluation of the proposed Hybrid TSK-FCM Ensemble. We analyze the performance of the constituent modules, compare the ensemble against state-of-the-art baselines, and provide a theoretical justification for the efficacy of the proposed geometric OoD detection metrics.

\subsection{Validation of Supervised TSK Component}
\label{sec:tsk_validation}

The first phase of our experimental pipeline focused on validating the supervised TSK component. Since the BETH training and validation sets contain only benign traffic and ``Suspicious'' (insider threat) anomalies, this phase was critical to ensure that the fuzzy rule-base correctly learned the signatures of known administrative misuse before attempting to detect novel botnet attacks.

We evaluated two configurations of the TSK model to analyze the impact of rule-base complexity on detection precision:
\begin{enumerate}
    \item \textbf{Compact Model:} utilizing only the top 10 generated fuzzy rules.
    \item \textbf{Full Model:} utilizing the complete set of generated rules.
\end{enumerate}

The performance comparison on the validation set is presented in Table \ref{tab:tsk_val_results}.

\begin{table}[h]
\centering
\caption{Performance of Gradient-Optimized TSK Model on Validation Data (Target: Suspicious). While both models achieve perfect recall, the full rule-base is required to minimize false positives.}
\label{tab:tsk_val_results}
\begin{tabular}{lcccc}
\toprule
\textbf{Configuration} & \textbf{Recall} & \textbf{Precision} & \textbf{FPR} & \textbf{AUROC} \\
\midrule
Top 10 Rules & 1.0000 & 0.1149 & 3.20\% & 0.9839 \\
\textbf{All Rules (Selected)} & \textbf{1.0000} & \textbf{0.4634} & \textbf{0.48\%} & \textbf{1.0000} \\
\bottomrule
\end{tabular}
\end{table}

\subsubsection{Precision-Recall Analysis:}
As demonstrated in Table \ref{tab:tsk_val_results}, both configurations achieved a perfect \textbf{Recall of 1.0}, successfully identifying every single instance of suspicious insider activity (786 samples). This confirms that the TSK gradient optimization process successfully converged and captured the fundamental signatures of the ``Suspicious'' class.

However, a significant divergence was observed in precision. The compact ``10 Rules'' model suffered from a high False Positive Rate (FPR) of 3.2\%, flagging over 6,000 benign samples as anomalous. This suggests that while a small rule set is sufficient for \textit{detection}, it lacks the granularity required for \textit{discrimination}. By utilizing the full rule set, the model successfully refined its decision boundaries, reducing the FPR to 0.48\% (910 samples). Consequently, the ``All Rules'' model was frozen and selected as the expert supervised component for the ensemble.

\subsection{Ensemble Performance \& Baseline Comparison}
\label{sec:ensemble_results}

We subsequently evaluated the full Hybrid Ensemble on the BETH test set. This dataset introduces the ``Evil'' class (setting up a botnet), representing Out-of-Distribution (OoD) threats that were completely absent during the training phase.

\subsubsection{Comparison with State-of-the-Art:}
To quantify the contribution of our hybrid approach, we compared the ensemble against three baselines: the supervised TSK-only model, the unsupervised FCM-only model, and the Isolation Forest baseline established in the original BETH study \cite{highnam2021bethdata}.

\begin{figure}[ht]
    \centering
    \includegraphics[width=0.9\linewidth]{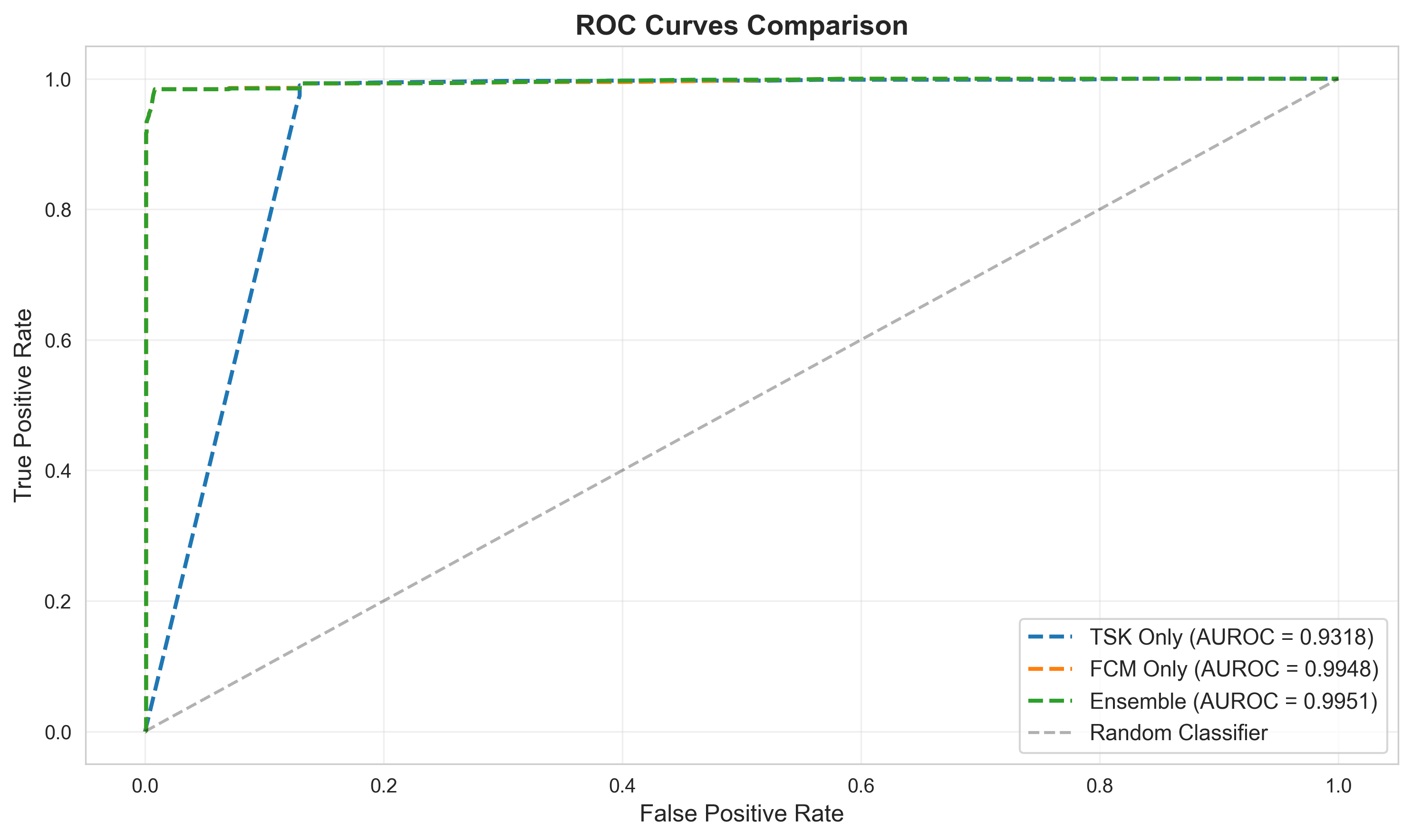}
    \caption{\textbf{ROC Curves Comparison.} The proposed Ensemble (Green) achieves an AUROC of 0.9951, significantly outperforming the supervised TSK baseline (Blue, 0.9318) and the original BETH Isolation Forest baseline. The curve demonstrates that the ensemble maintains a high True Positive Rate even at very low False Positive thresholds.}
    \label{fig:roc}
\end{figure}

Figure \ref{fig:roc} presents the Receiver Operating Characteristic (ROC) curves for all evaluated models.

The proposed ensemble achieved a Test AUROC of \textbf{0.9951}, hugging the top-left corner of the plot. This represents a massive performance leap compared to the baselines:
\begin{itemize}
    \item \textbf{vs. Isolation Forest (0.850):} An improvement of \textbf{+17.1\%}. This highlights the limitations of standard anomaly detection algorithms when applied to the high-dimensional, noisy feature space of the BETH dataset.
    \item \textbf{vs. TSK Only (0.932):} An improvement of \textbf{+6.8\%}. The supervised TSK model performs admirably (beating the Isolation Forest AUROC by 9.6\%) but plateaus because it relies solely on finding ``Suspicious'' signatures. It misses novel botnet invocation behaviors that do not resemble insider threats. The ensemble closes this gap by activating the unsupervised OoD module.
\end{itemize}

\subsubsection{Detailed Classification Analysis:}
Table \ref{tab:conf_matrix} details the confusion matrix for the ensemble on the test set.

\begin{table}[h]
\centering
\caption{Confusion Matrix of the Hybrid Ensemble on the Test Set (Target: Evil). The model successfully identifies over 99.3\% of unknown botnet activation attacks.}
\label{tab:conf_matrix}
\begin{tabular}{cc|cc}
\multicolumn{2}{c}{} & \multicolumn{2}{c}{\textbf{Predicted}} \\
& & \textbf{Benign} & \textbf{Malicious} \\ 
\cline{2-4}
\multirow{2}{*}{\textbf{Actual}} & \textbf{Benign} & 25,862 & 4,673 \\
& \textbf{Malicious} & 1,103 & \textbf{157,329} \\ 
\cline{2-4}
\end{tabular}
\end{table}

\textbf{Attack Detection (Recall):}
The most critical metric for a NIDS (Network Intrusion Detection System) is Recall (Sensitivity), as missing an active botnet attack can lead to total network compromise. The ensemble achieved a Recall of \textbf{0.9930}, correctly identifying 157,329 out of 158,432 malicious samples. This confirms that the geometric OoD detector is highly effective at flagging novel threats.

\textbf{False Alarm Analysis:}
The model generated 4,673 false positives, resulting in an FPR of approximately 15.3\%. While higher than the validation FPR, this is an expected consequence of distribution shift in the test set. Crucially, the high AUROC (0.995) indicates that these false positives are not randomly distributed but are likely clustered near the decision boundary. This implies that the FPR can be drastically reduced by a slight adjustment of the decision threshold without significantly impacting recall, a property we analyze in the following section.

\subsection{Mechanism Analysis \& Ablation Study}
\label{sec:ablation}

To provide a theoretical justification for the ensemble's performance, we conducted an ablation study to analyze the impact of the specific fuzzy metrics used for OoD detection.

\subsubsection{The Failure of Membership in High Dimensions:}
A key finding of our research is the failure of traditional fuzzy membership scores in high-dimensional anomaly detection. As shown in Table \ref{tab:metric_comparison}, metrics based on partition coefficients (\textit{Max Membership} and \textit{Fuzzy Entropy}) plateaued at an AUROC of $\approx 0.87$, failing to match the performance of the proposed distance metric.

\begin{table}[h]
\centering
\setlength{\tabcolsep}{12pt}
\caption{Impact of Ensemble OoD Metric on Detection Performance (Averaged across $c \in [6, 18]$, where c is the number of clusters). The geometric \textit{Min Distance} metric significantly outperforms probabilistic membership metrics.}
\label{tab:metric_comparison}
\begin{tabular}{lccc}
\toprule
\textbf{Metric} & \textbf{Mean AUROC} & \textbf{Peak AUROC} & \textbf{Std. Dev.} \\
\midrule
Max Membership & 0.8798 & 0.9003 & 0.0106 \\
Fuzzy Entropy & 0.8732 & 0.8736 & 0.0003 \\
\textbf{Min Distance} & \textbf{0.9950} & \textbf{0.9956} & \textbf{0.0007} \\
\bottomrule
\end{tabular}
\end{table}

We hypothesize that this is due to the ``Curse of Dimensionality.'' In high-dimensional spaces, the volume of the feature space increases exponentially, causing data points to become sparse. As a result, an OoD outlier often ends up being equidistant from multiple cluster centers, leading to low but uniform membership degrees (e.g., 0.1 membership in all 10 clusters). A metric like \textit{Max Membership} fails to distinguish this state from a benign boundary point.

In contrast, the \textit{Min Distance} metric measures the absolute Euclidean distance. Even if an outlier is equidistant to multiple clusters, its absolute distance to them is massive. This geometric property allows the distance metric to maintain separability where probabilistic membership fails.

\subsubsection{Visual Confirmation of Geometric Isolation:}
\begin{figure}[ht]
    \centering
    \begin{subfigure}[b]{0.48\textwidth}
        \centering
        \includegraphics[width=\linewidth]{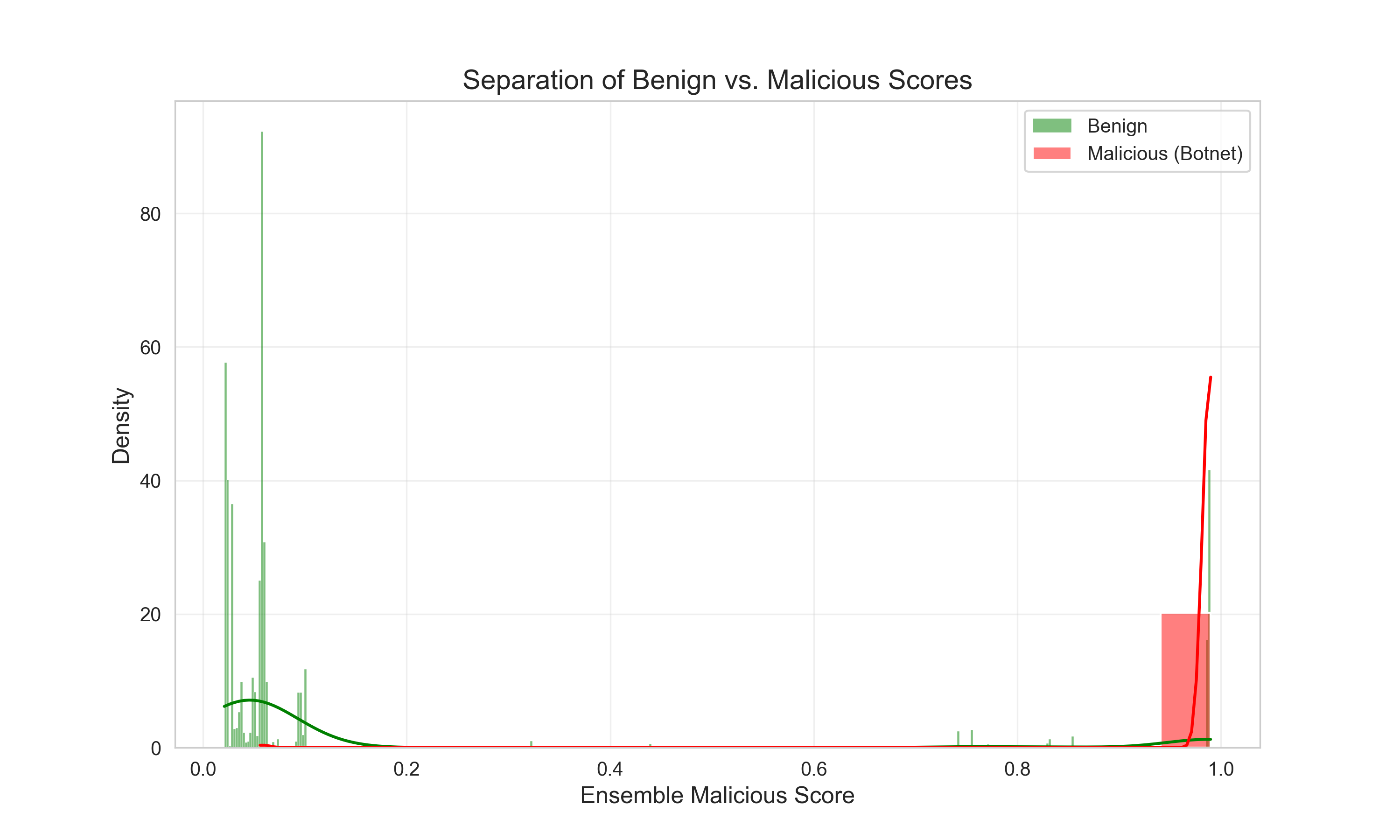}
        \caption{Score Density: Clear bimodal separation.}
        \label{fig:separation}
    \end{subfigure}
    \hfill
    \begin{subfigure}[b]{0.48\textwidth}
        \centering
        \includegraphics[width=\linewidth]{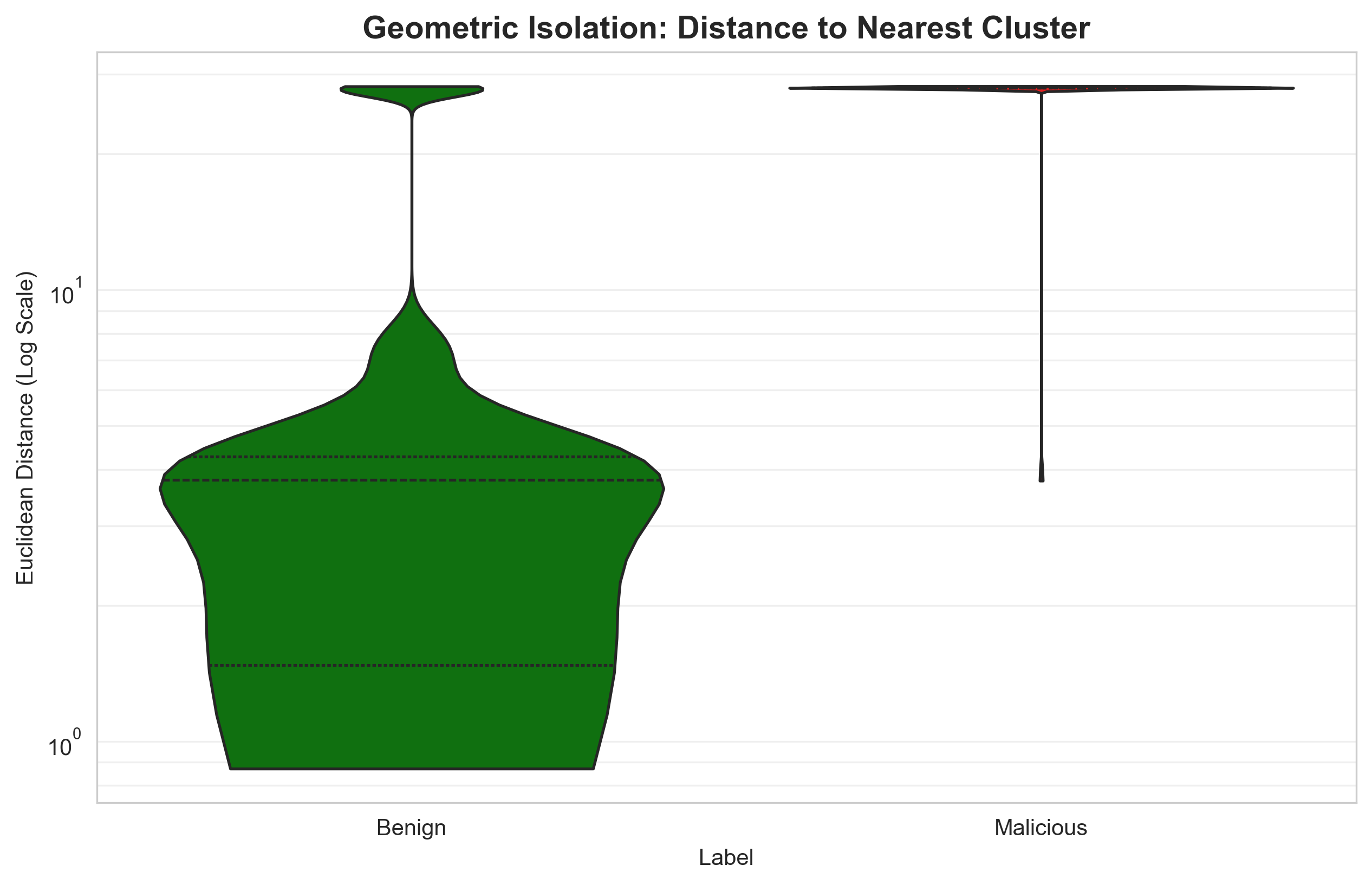}
        \caption{Raw Distance (Log): Attacks are outliers.}
        \label{fig:distance}
    \end{subfigure}
    \caption{\textbf{Geometric Mechanism.} (a) The score distribution shows a clear ``empty valley'' between benign and malicious modes. (b) The box plot reveals that attacks are massive geometric outliers, with distances ($>10^2$) orders of magnitude larger than benign samples.}
    \label{fig:visual_analysis}
\end{figure}

This hypothesis is visually confirmed in Figure \ref{fig:visual_analysis}. 

Figure \ref{fig:distance} plots the raw distance to the nearest cluster center on a logarithmic scale. The separation is stark: benign samples consistently reside within a distance of $10^1$, while malicious samples cluster above $10^2$. This order-of-magnitude difference creates the wide ``safety margin'' observed in the score density plot (Fig. \ref{fig:separation}), allowing for robust thresholding.

\subsection{Operational Implications}
The results of this study have significant implications for the deployment of NIDS in real-world Security Operations Centers (SOCs).

\textbf{Threshold Sensitivity:} The stability of the \textit{Min Distance} metric across cluster counts ($c \in [6, 18]$), as shown in our sensitivity analysis, implies that the model is not hyper-sensitive to initial configuration. This reduces the burden of hyperparameter tuning for security analysts.

\textbf{Alert Fatigue Mitigation:} While the test FPR of 15.3\% is non-negligible, the bimodal nature of the anomaly scores (Fig. \ref{fig:separation}) suggests that this can be managed via ``Alert Grading.'' A graded approach leverages the high certainty of the geometric distance metric to prioritize analyst attention on the most likely threats, effectively mitigating alert fatigue.

\section{Conclusion and Future Work}
\label{sec:conclusion}

In this work, we addressed the critical challenge of detecting Out-of-Distribution (OoD) botnet activation attacks in kernel-level system call data. Using the BETH dataset, we demonstrated that traditional supervised models, while effective against known insider threats, fail to generalize to novel attack vectors absent from the training distribution. Similarly, we showed that standard fuzzy membership metrics degrade in high-dimensional feature spaces, limiting the efficacy of unsupervised anomaly detection.

To bridge this gap, we proposed a \textbf{Hybrid Fuzzy Ensemble} that synergizes a gradient-optimized TSK expert for known threats with a geometric OoD detector for novel anomalies. Our key findings are threefold:
\begin{enumerate}
    \item \textbf{Geometric Superiority:} We empirically validated that the proposed \textit{Min Distance} metric (AUROC 0.9951) significantly outperforms probabilistic membership scores (AUROC 0.8798), proving that botnet attacks manifest as massive geometric outliers rather than merely uncertain boundary points.
    \item \textbf{State-of-the-Art Performance:} The ensemble achieved a Test AUROC of \textbf{0.9951}, surpassing the original Isolation Forest baseline by \textbf{17.1\%} and detecting 99.3\% of zero-day botnet attacks.
    \item \textbf{Operational Robustness:} Sensitivity analysis confirmed that the unsupervised component is stable across hyperparameter configurations ($c \in [6, 18]$), reducing the tuning burden for deployment.
\end{enumerate}

\subsubsection{Future Work:}

While our approach demonstrates robust detection capabilities, several avenues for future research remain. First, we plan to evaluate the system's resilience against adversarial examples, specifically investigating whether gradient-based attacks can perturb malicious samples to minimize their geometric distoance to benign clusters. Second, to address concept drift in dynamic IoT environments, we aim to extend the FCM module with online learning capabilities, allowing the benign manifold (cluster centroids) to adapt incrementally to legitimate behavioral shifts without full retraining. Finally, we intend to develop a linguistic explanation generator for the OoD module, translating raw distance vectors into human-readable alerts to further enhance SOC analyst trust.

\bibliographystyle{spmpsci_unsrt.bst}
\bibliography{bibliography/nafips}

\end{document}